\documentclass[pdflatex,sn-nature]{sn-jnl}

\usepackage{anyfontsize} 
\usepackage{graphicx}%
\usepackage{multirow}%
\usepackage{amsmath,amssymb,amsfonts}%
\usepackage{amsthm}%
\usepackage{mathrsfs}%
\usepackage[title]{appendix}%
\usepackage{xcolor}%
\usepackage{textcomp}%
\usepackage{manyfoot}%
\usepackage{booktabs}%
\usepackage{algorithm}%
\usepackage{algorithmicx}%
\usepackage{algpseudocode}%
\usepackage{listings}%
\usepackage{gensymb}

\theoremstyle{thmstyleone}%
\theoremstyle{thmstyletwo}%

\theoremstyle{thmstylethree}%

\usepackage{braket}
\usepackage{siunitx}
\newcommand{\um}{\si{\micro\metre}}

\usepackage{makecell}
\usepackage{ulem}
\usepackage[svgnames]{xcolor}
\newif\ifmydraft
\mydraftfalse 
\ifmydraft
    \usepackage[stamp=true, scale=1.5, color={[gray]{0.96} }]{draftwatermark}
\fi

\begin{document}

\title[A]{High fidelity CNOT gates in photonic integrated circuits using composite segmented directional couplers}


\author*[1]{\fnm{Jonatan} \sur{Piasetzky}}\email{piasetzky1@mail.tau.ac.il}

\author[1,2]{\fnm{Amit} \sur{Rotem}}
\author[1,2]{\fnm{Yuval} \sur{Warshavsky}}
\author[1,2]{\fnm{Yehonatan} \sur{Drori}}
\author[1]{\fnm{Khen} \sur{Cohen}}
\author[1,2]{\fnm{Yaron} \sur{Oz}}
\author[1,2]{\fnm{Haim} \sur{Suchowski}}

\affil[1]{\orgdiv{School of Physics and Astronomy}, \orgname{Tel Aviv University}, \orgaddress{\street{Chaim Levanon}, \city{Tel Aviv}, \postcode{6997801}, \country{Israel}}}

\affil[2]{\orgname{Quantum Pulse Ventures}, \country{Israel}}


\abstract{
Integrated photonic circuits are a promising platform for scalable quantum information processing, but their performance is often constrained by component sensitivity to fabrication imperfections. Directional couplers, which are crucial building blocks for integrated quantum logic gates, are particularly prone to such limitations, with strong dependence on geometric and spectral parameters that reduce gate fidelity. Here, we demonstrate that composite segmented directional couplers (CSDC) offer a fabrication-tolerant alternative that enhances gate fidelity without active tuning. We design and fabricate a fully integrated photonic controlled-NOT (CNOT) gate using both uniform and composite coupler variants and compare their performance via simulation, classical characterization, and quantum two-photon interference. The composite design reduces the average error probability by nearly a factor of two and variability by a factor of five. The residual error is primarily limited by photon indistinguishability. Classical matrix reconstruction confirms improved agreement with the ideal CNOT operation. These results establish CSDCs as compact, passive, and foundry-compatible building blocks for robust, scalable quantum photonic circuits.}
\maketitle

\section*{Main}\label{sec:Main}
Quantum computing offers exponential speedups for specific problems such as integer factorization \cite{shor_algorithms_1994}, quantum simulation~\cite{montanaro_quantum_2016,harris_quantum_2017}, and certain optimization and search tasks~\cite{grover_quantum_1997}. Among the leading platforms for quantum information processing, photonic systems are especially attractive due to their low decoherence and room-temperature operation~\cite{arrazola_quantum_2021,rudolph_why_2017, bourassa_blueprint_2020}. In linear optical quantum computing (LOQC), quantum gates are implemented using beam splitters, phase shifters, single-photon sources, and detectors, with entangling operations realized via measurement-induced nonlinearity~\cite{knill_scheme_2001, kok_linear_2007}. Although LOQC is, in principle, scalable using cluster-state or fusion-based architectures~\cite{alexander_manufacturable_2024,Gimeno-Segovia2015,gimeno-segovia_towards_2016}, its performance critically depends on the precision of individual optical components, particularly beam splitters and interferometers, which motivates the development of more robust and fabrication-tolerant photonic building blocks.

Integrated photonics is rapidly emerging as a scalable and practical platform for quantum information processing, offering compact device footprints, low optical losses, and compatibility with mature semiconductor fabrication processes. Quantum photonic circuits impose stringent requirements on component performance. As quantum photonic circuits scale in complexity, their performance becomes increasingly affected by the performance of the individual building blocks and the requirement for reliable passive components increases. 

Directional couplers, which enable coherent interference and state transfer between adjacent waveguides, are fundamental to these circuits, used as single-qubit gates and as parts of interferometers and multi-qubit operations such as the controlled-NOT (CNOT) gate. However, conventional directional couplers exhibit strong dependence on wavelength, waveguide geometry, coupling length and refractive index, rendering them highly susceptible to fabrication imperfections and thermal variations. This sensitivity degrades gate fidelity, particularly in post-selected LOQC schemes, where high-visibility quantum interference is essential for entanglement and logic correctness. In multi-qubit gates like the CNOT, even small deviations in splitting ratio or accumulated phase can lead to incorrect output states or loss of entanglement fidelity.

To address these challenges, a new class of devices known as composite segmented directional couplers (CSDCs) has been proposed~\cite{Kyoseva2019,ivanov_high-fidelity_2022} and recently demonstrated using classical light~\cite{katzman_robust_2022, kaplan_segmented_2022}. Inspired by composite pulse protocols in nuclear magnetic resonance, CSDCs divide the coupling region into multiple segments with engineered detunings and coupling rates, unlike conventional directional couplers, which have a uniform cross-section in the interaction region, hereafter referred to as uniform directional couplers. This segmented structure allows for destructive interference of systematic errors, effectively averaging out imperfections over the full device length. Unlike Mach–Zehnder-based or adiabatic designs~\cite{ramadan_adiabatic_1998}, CSDCs provide robustness passively with no need for active tuning, and with only a modest increase in device footprint. Previous work has shown that CSDCs exhibit significantly flatter spectral response~\cite{katzman_robust_2022} and reduced sensitivity to fabrication deviations compared to uniform directional couplers~\cite{katzman_robust_2022,kaplan_segmented_2022,cohen_robust_2025}, but their applicability to quantum operations has remained untested.

Here, we report the first implementation of a photonic two-qubit CNOT gate entirely based on CSDCs. We extend the CSDC framework to the multi-qubit quantum regime by realising a fully integrated CNOT gate that replaces every directional coupler with its CSDC counterpart. Through numerical simulation, classical characterization, and quantum two-photon experiments, we show that CSDCs outperform uniform couplers in fidelity, robustness, and reproducibility. We achieve CNOT gates with a mean error probability of $3.01\% \pm 0.47\%$, compared to $5.5\% \pm 2.1\%$, for the legacy uniform design. These results establish CSDCs as the leading passive, fabrication-tolerant option for scalable quantum photonics, surpassing the prevailing uniform architecture. To our knowledge, this is the first quantum two-qubit gate built on CSDCs, demonstrating that composite design principles seamlessly transfer from classical optics to quantum logic and providing a critical building block for resilient, fault-tolerant photonic processors.

\section*{Design and Simulation}
\begin{figure}[H]
    \centering
    \includegraphics[width=\linewidth]{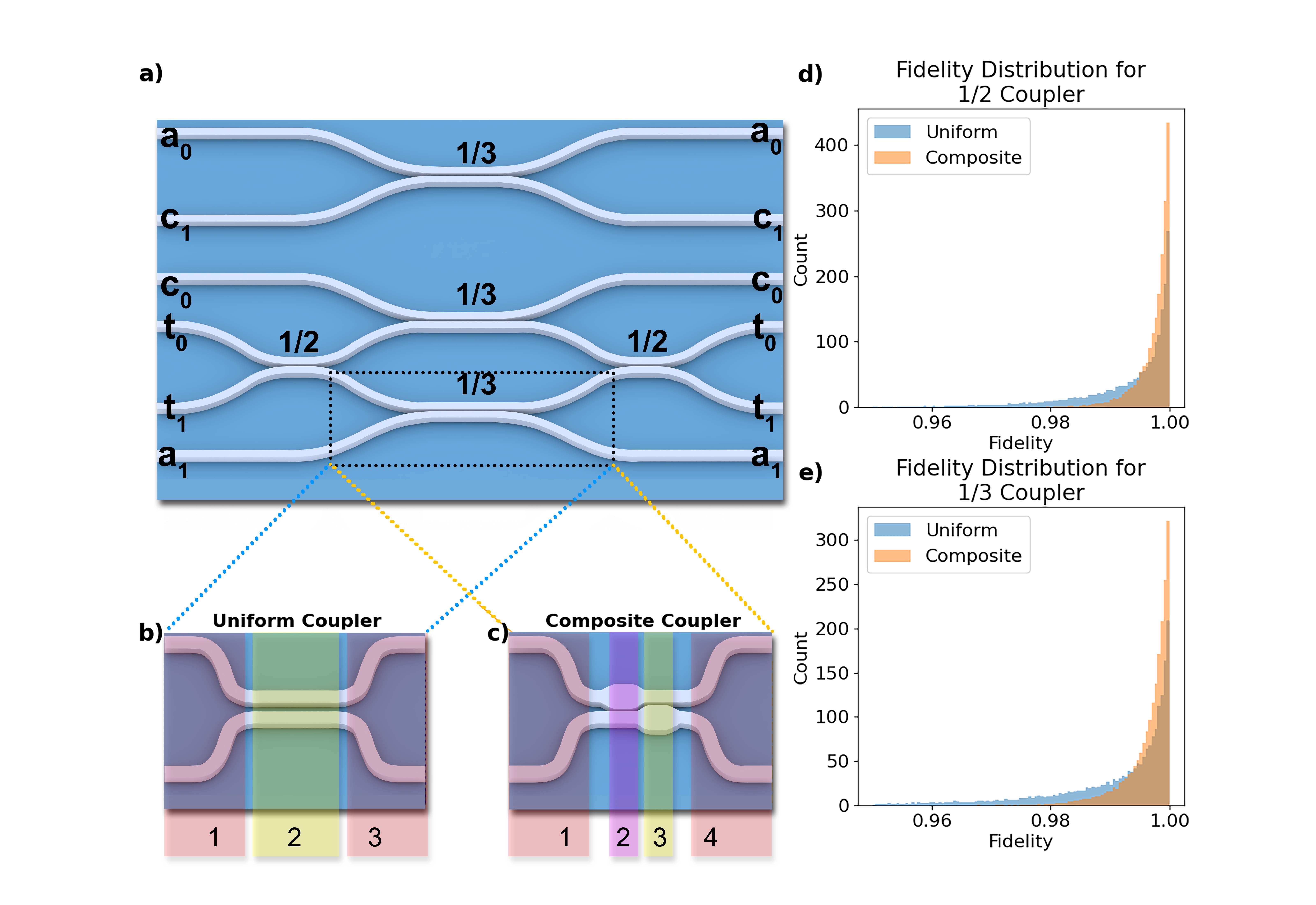}
    \caption{
    \textbf{Design and robustness comparison of composite and uniform directional couplers in an integrated CNOT gate.}
    \textbf{a)} Schematic layout of the post-selected linear optical CNOT gate, consisting of directional couplers with $1/3$ and $1/2$ splitting ratios. The architecture requires two $1/2$ and three $2/3$ directional couplers.
    \textbf{b)} Zoom-in of a schematic conventional uniform directional coupler with a fixed cross-section. The ingoing and outgoing S-bend areas are marked as areas 1 and 3 respectively. The yellow area marked as 2 is the uniform cross-section interaction region.
    \textbf{c)} Zoom-in of a schematic composite segment directional coupler, composed of multiple segments with varying waveguide widths to implement detuned coupling. The different segments are marked as 2 and 3, and have a 2 $\mu m$ linear taper between them.
    \textbf{d)} Simulated gate error probability distribution function for $1/2$ directional couplers. This Monte Carlo simulation compares uniform (blue) and composite (orange) designs with fabrication errors distributions that follow our previous characterization, discussed in the main text.
    \textbf{e)} Same as (d), for $1/3$ directional couplers. In both cases, the composite design exhibits significantly reduced sensitivity to fabrication-induced variations in waveguide width.
    }
    \label{fig:fig1_cnot_arch}
\end{figure}

We implemented a fully integrated linear optical CNOT gate based on the dual rail architecture of Ref.~\cite{ralph_linear_2002}, previously demonstrated in bulk~\cite{obrien_demonstration_2003} and integrated platforms~\cite{politi_silica--silicon_2008}. The architecture (Fig. \ref{fig:fig1_cnot_arch}\textbf{a}) follows a post-selected design, where nonclassical interference and measurement-induced collapse enable entangling two-qubit operations, using directional couplers to perform the beam-splitting and merging functions necessary for gate operation. In the dual rail path encoding, each qubit is composed of two different spatial modes, one for the logical $\ket{0}$ and another for the logical $\ket{1}$. Thus, our integrated three qubit CNOT is composed of six waveguides, as shown in Fig. \ref{fig:fig1_cnot_arch}\textbf{a}, titled $c_0, c_1, t_0, t_1, a_0, a_1$ for the $\ket{0}$ and $\ket{1}$ modes of the control, target and ancila qubit respectively. An ideal CNOT opeation flips the target qubit if and only if the control qubit is in the logical $\ket{1}$ state. The CNOT design comprises five directional couplers - two with a reflectivity of 1/2 and three with a reflectivity of 1/3. These couplers transform input modes via unitary operations $U_R = \exp(i \sigma_x \phi_R)$, where $\sigma_x$ is the Pauli-X operator and $\phi_R=\arccos{\left(\sqrt{R}\right)}$ is the coupling angle of a coupler with reflectivity R.

In the following, we compare the uniform designs, consisting of a single segment (Fig. \ref{fig:fig1_cnot_arch}\textbf{b}) and the CSDCs design consisting of four fixed-width segments connected by linear tapers, each engineered for optimal detuning and coupling (For brevity, Fig. \ref{fig:fig1_cnot_arch}\textbf{c} shows a schematic two segment design). 
The CSDC designs were developed using a loss-weighted fidelity optimization process, which considered fabrication variability. Details of the optimization process are provided in the Methods section, and the final dimensions that were eventually fabricated are summarized in Table~\ref{tab:design}. 

The isolated performance of both the 1/2 and 1/3 single couplers were evaluated. The results comparing uniform and CSDC designs are shown in Fig. \ref{fig:fig1_cnot_arch}\textbf{d-e}, and display improved performance for the CSDC designs. While the CSDC designs have mean fidelities of $99.74\% \pm 0.32\%$ and $99.64\% \pm 0.42\%$, for the 1/2 and 1/3 couplers respectively, the uniform designs have mean fidelities of $99.2\% \pm 1.1\%$ and $98.9\% \pm 1.4\%$ for the same couplers.

\begin{table}[h]
    \centering
    \begin{tabular}{|c|c|c|c|c|}
    \hline
    & Uniform 1/2 & CSDC 1/2 & Uniform 1/3 & CSDC 1/3 \\
    \hline
    \makecell{Waveguide 1\\ widths [nm]} & [500] & [510, 472, 528, 490] & [500] & [496, 528, 472, 504] \\ 
    \hline
    \makecell{Waveguide 2\\ widths [nm]} & [500] & [490, 528, 472, 510] & [500] & [504, 472, 528, 496] \\
    \hline
    Segment lengths [nm]  & [10029] & [26, 48, 111, 138] & [13243] & [1696, 103, 103, 1696] \\
    \hline
    Total length [nm] & 10029 & 10323 & 13243 & 13598 \\
    \hline
    \end{tabular}
    \caption{Design parameters for uniform and composite directional couplers, as received by the optimization process. These parameters were ultimately fabricated for the experimental validations. The composite segment lengths do not sum up to the total length due to the existence of a 2 $\mu m$ linear taper between each segment.}
    \label{tab:design}
\end{table}

\section*{Results}\label{sec:results}
The photonic circuits were fabricated by Applied NanoTools Inc. \cite{noauthor_nanosoi_nodate}, a commercial fabrication service. Uniform and CSDC CNOT variants were fabricated side by side for direct comparison. Five copies of each variant were characterized with quantum measurements, as detailed in the Methods section. Two-photon interference visibility was measured for the input states \(C_0T_0\) and \(C_0T_1\), resulting in 98\% visibility, confirming high photon indistinguishability and minimal polarization drift in the chip. This also sets an upper bound on achievable fidelity and confirms that residual errors and CSDC CNOT performance are primarily source-limited (Ref. ~\cite{nakav_quantum_2025}, Eq.(5)). Linking the measured HOM visibility to the isolated single coupler characterization, we modeled and simulated the full CNOT circuit and extracted the expected error probability. 

\begin{figure}[h]
    \centering
    \includegraphics[width=\linewidth]{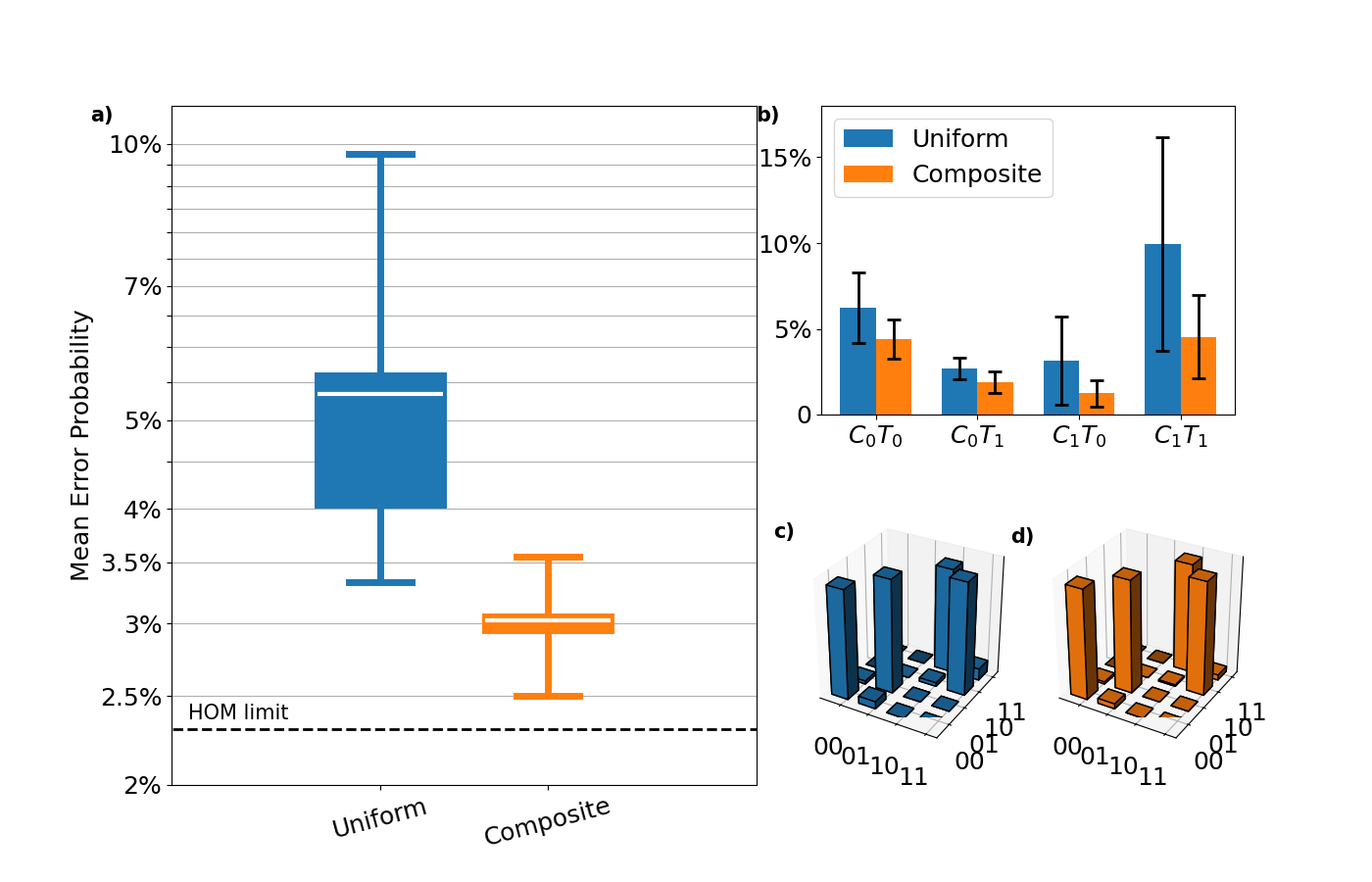}
    \caption{\textbf{Quantum validation of CNOT gates implemented with uniform and composite
segment directional couplers}. \textbf{a)} The mean error probability, over all input basis states, for CNOT gates implemented with uniform ({\color{blue}blue}) and composite ({\color{orange}orange}) directional couplers. The limit of best fidelity according to indistinguishably is shown in a dashed line (more information on that in the main text).
\textbf{b)} Error probability measured for each of the four computational
basis inputs: $C_0T_0$, $C_0T_1$, $C_1T_0$, and $C_1T_1$. Error bars represent one standard deviation from repeated
measurements. \textbf{c-d)} The probability detection matrix ofr uniform (c) and composite (d). Each bar height represents the normalized probability of detecting a coincidence in an output state for a given input state. The composite implementation shows improved agreement with the expected ideal CNOT truth table. In total, 5 CNOTs of each kind were measured.}
    \label{fig:fig2_cnot_fids}
\end{figure}

Next, we injected the four computational basis states (\(C_0T_0\), \(C_0T_1\), \(C_1T_0\), \(C_1T_1\)) and measured coincidence rates to extract the error probability. For the uniform design, we measured a mean error probability of 5.5\% ± 2.1\%, in excellent agreement with the simulated 5.4\% ± 3.0\%. The CSDC design delivered an experimental mean error probability of 3.01\% ± 0.47\%, outperforming the simulated 4.0\% ± 1.1\%. As shown in Fig.~\ref{fig:fig2_cnot_fids}, and summarized in Table \ref{tab:results}, the composite approach reduces the error probability by approximately \(45\%\) and suppresses the standard deviation by a factor of 4.6, validating that the advantages of CSDCs persist from single-gate operation to more complex multi-gate quantum circuits. 

\begin{table}[h]
    \centering
    \begin{tabular}{|c|c|c|}
    \hline
    & Experiment & Simulation \\
    \hline
    Uniform & 5.5\% ± 2.1\% & 5.4\% ± 3.0\%  \\ 
    \hline
    Composite & 3.01\% ± 0.47\% & 4.0\% ± 1.1\%  \\
    \hline
    HOM limit & \multicolumn{2}{c|}{2.3\%} \\
    \hline
    \end{tabular}
    \caption{Main results comparing CNOT circuits of CSDC and uniform design, for both experiment and simulation. Simulation results rely on single coupler experimental measurements and HOM visibility. The HOM lower error limit is calculated according to~\cite{nakav_quantum_2025}, Eq.(5) and assumes perfect directional couplers.}
    \label{tab:results}
\end{table}



\begin{figure}[h]
    \centering
    \includegraphics[width=\linewidth]{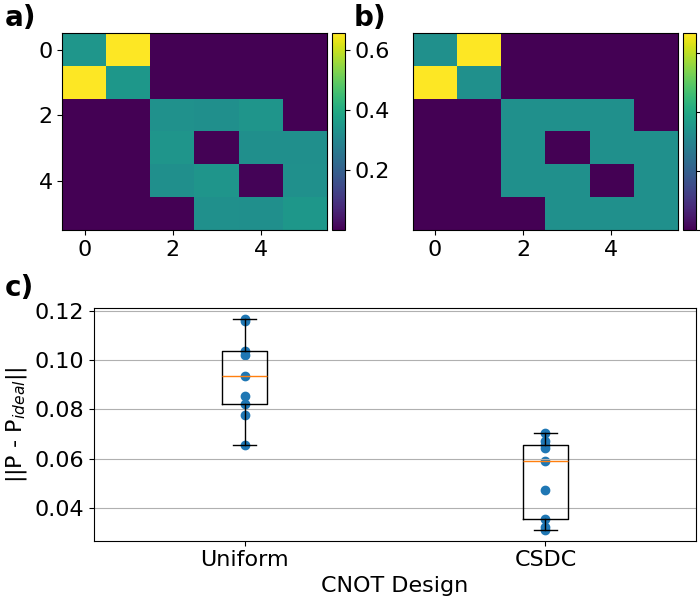}
    \caption{
    \textbf{Classical validation of CNOT gate fidelity using power-only characterization.}
    \textbf{a)} Power of the experimentally reconstructed transformation matrix for a representative uniform CNOT gate, after applying the Sinkhorn-based decomposition to isolate the contribution of input/output couplers. 
    \textbf{b)} Corresponding power transformation matrix for the ideal CNOT operation.
    \textbf{c)} Frobenius distance between reconstructed ($P$) and ideal CNOT ($P_{ideal}$) power matrices for two designs, uniform and CSDC. Each design had nine copies. CSDC design shows smaller distance from ideal matrix, indicating improved fidelity and robustness under identical fabrication conditions.}
    \label{fig:fig3_classical}
\end{figure}

Prior to quantum testing, we performed classical optical characterization of the fabricated circuits. Since the CNOT circuit is composed exclusively of linear optical components, and we neglect loss for the purposes of this analysis, we model the device as a generic 6×6 unitary linear transformation acting on the waveguide modes. We characterized the power response of this transformation using the Sinkhorn decomposition-based algorithm described in Ref. \cite{hoch_characterization_2023}, which isolates and compensates for the lossy and noisy contributions of the optical input/output grating couplers (Fig. \ref{fig:fig3_classical}\textbf{a}).

We compared the reconstructed matrix amplitude to that of the ideal CNOT transformation (Fig. \ref{fig:fig3_classical}\textbf{b}) by calculating the Frobenius distance between them. This procedure was repeated for nine copies of each CNOT circuit variant, CSDC and uniform. The Frobenius distance between experimental and ideal matrices was smaller for CSDC circuits (Fig. \ref{fig:fig3_classical}\textbf{c}), indicating superior performance under the given fabrication variations and measurement conditions.

\section*{Discussion}
We have integrated composite segmented directional couplers into a two-qubit photonic CNOT gate and verified performance gains over uniform couplers through classical matrix reconstruction and two-photon tests. The composite design lowers error probability and suppresses variability without a need for active tuning, an appreciable footprint increase, or additional control overhead, and is compatible with standard foundry constraints. These results validate a practical, passive path to high-fidelity, scalable quantum photonic circuits and provide a compact building block for resource-efficient, ultimately fault-tolerant photonic computing. We achieved nearly a factor-of-two reduction in error probability down to $3.01\% \pm 0.45\%$, with the majority of this error attributed to the photon distinguishability of our imperfect source, which showed approximately 98\% visibility in HOM interference. This error is unaffected by the CSDC design and therefore requires improvements in the source itself. Our simulations and analysis predict that with a perfectly indistinguishable photon-pair source, the CNOT error probability should decrease to roughly 1\%.

These benefits extend to quantum architectures reliant on entanglement resource generation, such as fusion-based and cluster-state schemes. Reduced device error translates directly to higher success probabilities and reduced entanglement loss. Furthermore, lower baseline gate error and narrower statistical spread move hardware closer to error-correction threshold regimes, typically in the 1–2\% range, where error-correction overheads shrink markedly. In practical terms, robust passive couplers can reduce the number of physical modes and entangling steps required to realize a target logical error rate and may enable below-threshold performance for select photonic encodings. Moreover, the fivefold reduction in variability implies improved stability and reproducibility, making the CSDC design especially attractive for wafer-scale foundry fabrication and suggesting a higher production yield. We expect CSDCs to play a foundational role in scalable photonic quantum processors.

\bibliography{references}

\section*{Methods}

\subsection*{Simulation and Modeling}
Both uniform and CSDC designs were modeled using silicon-on-insulator (SOI) waveguides at 1550 nm with a 220 nm thickness and a 650 nm center-to-center gap. Coupled-mode theory simulations, performed using Lumerical and the effective trench medium model~\cite{warshavsky_accurate_2025}, yielded spatial profiles of coupling, $\kappa(z)$, and detuning, $\Delta\beta(z)$. The resulting unitary was obtained by integrating the Hamiltonian over the entire interaction length, $L$. We also account for the S-bend regions, in which the waveguides are brought close together or further apart, by incorporating a pre-characterized S-bend transfer matrix to recover the overall unitary transformation:
\begin{equation}
U_{\text{DC}} = U_{\text{S-bend}} \cdot\mathcal{T}\left[\exp \left[-i \int_0^L\left(\begin{array}{cc}
    -\Delta\beta\left(z\right) & \kappa^*\left(z\right) \\
    \kappa\left(z\right) & \Delta\beta\left(z\right)
    \end{array}\right)\right] d z\right] \cdot U^T_{\text{S-bend}}
\end{equation}
where $\mathcal{T}$ is the time ordering operator, and we notice that due to symmetry the unitary transformation of the outgoing S-bend region is the transpose of the ingoing S-bend region.

Waveguide parameters (width, gap, height) were swept in Lumerical FDE and FDTD simulations, using effective trench model ~\cite{warshavsky_accurate_2025} to extract coupling and detuning values. To reduce back-scattering loss, transitions between segments were designed with a 2 $\mu m$ linear taper, assumed to be adiabatic \cite{fu_efficient_2014}. The S-bend matrix was applied to complete the model, it was simulated using Lumerical's FDTD simulation; The S-bend is a \(L \approx 12.3\;\um\) long Euler Bend taking the rails from a (center to center) gap of 5 \um, to 0.65 \um, at a constant waveguide width of 0.5 \um.

\subsection*{Design Optimization}
Design optimization minimized a loss-weighted fidelity metric under fabrication variability assumptions. The variability was modeled as a normally distributed 2 nm common-width error, i.e., $\Delta w_1 = \Delta w_2 = \Delta w \sim \mathcal{N}(0, 2\text{ nm})$, which mostly matches previous characterization performed by us. 

Optimized parameters minimize the SU(n) trace fidelity~\cite{cohen_robust_2025,pedersen_fidelity_2007} error with a loss correction term:
\[\text{Cost} = 1 - |\text{Tr}(T^\dagger U)/n|^2 \times 10^{-0.1\alpha L}\]
where $T$ is the target gate, $U$ is the simulated unitary and $\alpha$ is the loss-per-length coefficient of the waveguides. Fabrication error was modeled as 2 nm Gaussian variation and $\alpha$ was taken to be 1.5 dB/cm, as characterized by Applied Nanotools Inc. \cite{noauthor_nanosoi_nodate}, the fabrication vendor. The optimization was implemented in Julia, using precomputed Hamiltonian grids.
\subsection*{Device Fabrication}
The photonic circuits were fabricated by Applied NanoTools Inc. \cite{noauthor_nanosoi_nodate}, a commercial fabrication service. Fabrication was performed as part of a Multi-Project Wafer run, using a standard 220 nm SOI wafer. The designs were patterned using electron-beam lithography, followed by reactive ion etching to define the photonic layer. All waveguides were fully etched down to the buried silica layer. Lastly, a layer of 2.2 \um\ of oxide was deposited post-etching. 

\subsection*{Quantum Characterization}
\begin{figure}
    \centering
    \includegraphics[width=1\linewidth]{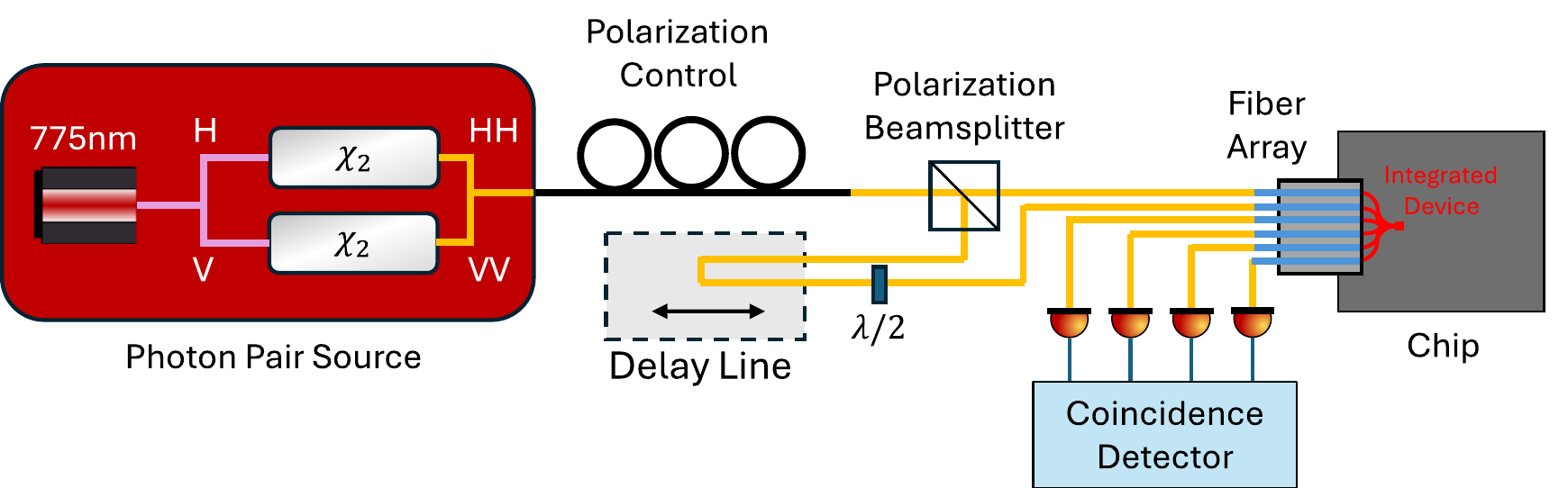}
    \caption{The CNOT experimental setup. Photons from a commercial  SPDC source are separated and coupled simultaneously to the input ports of an integrated CNOT gate. The four outputs ports of the CNOT gates are measured using 4 cryogenic photon counters, and coincidences are extracted using a time tagger.}
    \label{fig:cnotsetup}
\end{figure}

The experimental setup is shown in Fig. \ref{fig:cnotsetup}. Our photon pair source (OzOptics Ruby) utilizes type-0 SPDC process in periodically poled lithium niobate waveguide and tuned to generate degenerate photon pairs at 1550 nm. The signal and idler were ejected to the same polarization-maintaining (PM) fiber and separated using a polarization controller (PC) and a polarizing beam splitter. One arm was coupled to a motorized delay stage and a half-wave plate, which were used to maximize the HOM interference visibility. Both arms were directed into separate PM fibers, and each photon was independently coupled into one of the qubit inputs of the photonic CNOT chip.

The output state was collected using a matched PM fiber array and directed to a set of superconducting nanowire single-photon detectors (SNSPDs) manufactured by Quantum Opus. Coincidence detection was performed using a TimeTagger 20 (Swabian Instruments), with a coincidence window of 500 ps Each data point was acquired over 4000 seconds. Non-interfering outputs yielded approximately one coincidence per second on average.

Input and output coupling is achieved by aligning the Gaussian fiber mode to the on-chip guided mode using a grating coupler array designed for 1550 nm and an 8$\degree$ incidence angle.

\subsection*{Classical Characterization}
Classical characterizaiton was performed with a similar setup, the light source used was a broadband erbium-doped fiber amplifier source. The output spectra were measured using a Yokogawa Optical Spectrum Analyzer (OSA). Similarly to the quantum experiment, PM fibers and PCs were used to ensure polarization mode alignment at all inputs. Light was injected into and collected from the chip via a grating coupler array, aligned to a matching fiber array.

\subsection*{Single-Coupler Noise Characterization}
Since it was difficult to perform full state tomography with our system, we evaluated the single coupler performance by fabricating many ($\sim 150$) isolated directional couplers. Classical light was used to extract the splitting ratios, and the distribution of the measured splitting ratio was used to approximate error distribution using the following noisy gate model:
$$\tilde{U_i} = \exp \left(i \phi_i \left[\left(X_i+\epsilon_x\right)\sigma_x + \left(Z_i + \epsilon_z\right)\sigma_z\right]\right)$$
\begin{table}
    \centering
    \begin{tabular}{|l|*{2}{c|}}
        \hline
         & \(\epsilon_x\) & \(\epsilon_z\) \\
        \hline
        Uniform    & 0.022(1) & 0.125(10) \\
        \hline
        Composite  & 0.027(4) & 0.066(6) \\
        \hline
    \end{tabular}
    \caption{Fitted error parameters from the noisy gate model}
    \label{tab:error_params}
\end{table}
Where $\phi_i, X_i, Z_i$ describe the nominal zero error gate fabricated. The fitted values of $\epsilon_x$ and $\epsilon_z$ are shown in Table \ref{tab:error_params}, and were used to evaluate the single coupler performance and to simulate and predict the CNOT performance.
\section*{Competing Interests}
The authors declare no competing interests.

\section*{Data Availability}
Data supporting the findings are available upon request.

\section*{Acknowledgements}
We thank Applied Nanotools Inc.\cite{noauthor_nanosoi_nodate} for fabrication support.


\end{document}